\documentclass[prc,showpacs,superscriptaddress]{revtex4}
\usepackage{graphicx}
\usepackage{psfig}
\usepackage{epsfig}
\usepackage{bm}
\usepackage{amsmath}

\begin{document}

\title{Pairing properties of nucleonic matter employing dressed nucleons}

\author{H. M\"{u}ther}
\affiliation{Institut f\"ur Theoretische Physik, Universit\"at T\"ubingen,\\
D-72076 T\"ubingen, Germany}

\author{W. H. Dickhoff}
\affiliation{Department of Physics, Washington University, St. Louis, 
Missouri 63130, USA}

\date{\today}

\begin{abstract}
A  survey of pairing properties of nucleonic matter is presented that includes
the off-shell propagation associated with short-range and tensor correlations.
For this purpose, the gap equation has been solved in its most general form
employing the complete energy and momentum dependence of the normal 
self-energy contributions. The latter correlations include the self-consistent
calculation of the nucleon self-energy that is generated by the summation of
ladder diagrams. This treatment preserves the conservation of particle number
unlike approaches in which the self-energy is based on the
Brueckner-Hartree-Fock approximation. A huge reduction in the strength as well
as temperature and density range of  ${}^3S_1$-${}^3D_1$ pairing is obtained
for nuclear matter as compared to the  standard BCS treatment. Similar dramatic
results pertain to ${}^1S_0$ pairing of neutrons in neutron  matter. 
\end{abstract}


\maketitle

\section{INTRODUCTION}

Advances in the understanding of the single-particle (sp) properties of 
nucleons in nuclei and nuclear matter~\cite{diba04} demonstrate the dominant
influence of short-range and tensor correlations in generating the distribution
of the spectral strength.
One aspect of this influence is expressed in the global depletion of Fermi
sea due to these correlations.
A recent experiment from NIKHEF puts this depletion of the proton Fermi sea
in ${}^{208}$Pb at a little less than 20\%~\cite{bat01} in accordance with
earlier nuclear matter calculations~\cite{vond1}.
Another consequence of the presence of short-range and tensor correlations
is the appearance of high-momentum components in the ground state to compensate
for the depleted strength of the mean field.
Recent JLab experiments~\cite{rohe:04} indicate that the amount and location 
of this strength is consistent with earlier predictions for finite 
nuclei~\cite{mudi:94} and calculations for infinite 
matter~\cite{frmu:03,thesfr}.
Of particular relevance is the observed energy distribution of the sp strength
of the high-momentum nucleons that is located at energies far removed from
the Fermi energy.
This situation is similar to the case of nuclear matter~\cite{vond2} and
also holds above the Fermi energy.
Such sp strength distributions lead to substantial modifications of the
calculated properties of low-energy phenomena, like pairing, as compared to  
those generated by a traditional mean-field treatment. 

Pairing properties of nuclear and neutron matter have been studied for
quite some time. A recent review can be found in Ref.~\cite{dehj:03}.
The study of pairing correlations in infinite quantum systems is an interesting
question in general relating nuclear and neutron matter to the classical systems
studied in condensed matter physics as well as to the studies of pairing
correlations in fermionic atomic gases~\cite{qijn:05}. The study of pairing
correlations in nuclear and neutron matter is of particular interest for the
understanding of properties of neutron stars. The formation of a BCS gap has a
significant effect on the neutrino emissivity, which is crucial
for the cooling of neutron stars~\cite{bla:05}. The existence of superfluid
layers will affect the rotation of neutron stars.
  
Of particular interest is the observation that several calculations for the
bare nucleon-nucleon (NN) 
interaction in the ${}^3S_1$-${}^3D_1$ coupled channel 
lead to a sizable gap of around 10 MeV at normal nuclear matter 
density~\cite{ars:90,vgdpr:91,bbl:92,tt:93,bls:95}. Those calculations
typically solve the BCS gap equation using a mean-field sp
propagator with sp energies as determined \textit{e.g.} in a
Brueckner-Hartree-Fock (BHF) calculation.
  
The empirical data of finite nuclei do not exhibit any indications for
such strong proton-neutron pairing correlations, which would correspond to a 
gap as large as 10 MeV. Therefore one may conclude that such evaluations of a 
pairing gap in infinite matter yield quite different results from those
observed for finite nuclei. This appears plausible since 
pairing effects are very
sensitive to the sp spectrum close to the Fermi energy, which is 
continuous in infinite matter while it is discrete in
finite nuclei.

This observation may also cast some doubt on the reliability of the approach to
determine the pairing gap for infinite matter as outlined above. Therefore
attempts have been made to go beyond this mean-field approach. One issue
discussed in the literature has been the role of vertex corrections, 
\textit{i.e.} the
medium dependence of the NN interaction to be used in solving the gap
equation~\cite{ZLS:02,sedr:03,sls:03,sls:05}. Using effective interactions like
the Gogny force~\cite{gogny} Shen \textit{et al.}~\cite{sls:03,sls:05} 
observe indeed a
significant effect, which is larger for nuclear matter than for neutron
matter. The effects of the so-called induced interaction could indeed affect
the low-energy spectroscopy of nuclear matter in a significant way. Before
conclusions can be drawn, calculations employing realistic interactions should
be performed, which, unfortunately, are very difficult~\cite{induced}. 
In addition, it is possible that polarization effects are different in
nuclear matter and finite nuclei on account of the difference of the sp
spectra discussed above.
 
In the present work we would like to focus the attention to the study of 
pairing
correlations with a proper treatment of the nucleon propagator that accounts
for the effects of short-range and tensor correlations.  
The crucial effects of short-range and tensor correlations on the
distribution of the sp strength have recently been treated at a very 
sophisticated level. Various groups have developed techniques to evaluate the 
sp strength from realistic NN interactions, like the Argonne V18~\cite{arv18} 
or the CDBonn~\cite{cdbonn} interaction, within the self-consistent Green's 
function (SCGF) approach~\cite{rd,dwvnw,bozek0,bozek1,bozek2,dewulf:03,frmu:03,diep:03,frmu:05}. 
Such calculations reproduce the energy- and momentum-distribution of the sp 
strength corresponding to high-momentum nucleons as observed in 
experiments~\cite{rohe:04} and account for the depletion of the mean-field
strength obtained in~\cite{bat01}.

In these calculations the scattering equations for two nucleons in the
medium are solved employing dressed sp propagators that contain
the complete information about the energy- and momentum-distribution of 
the sp strength. 
The resulting scattering matrix $T$ is then used to calculate the nucleon
self-energy. The solution of the Dyson equation, employing 
this complex and energy
dependent self-energy, yields the sp propagator that enters the
$T$ matrix equation to close the self-consistency cycle. 
In determining the two-nucleon Green's function or the corresponding scattering
matrix $T$, one has to face the problem of the so-called pairing instabilities
that reflect the existence of NN bound state solutions in the scattering
equation.
The presence of such solutions provides a major numerical obstacle for a 
self-consistent evaluation of one- and
two-body propagator within the normal Green's function approach
at densities where such instabilities can occur. 
For this reason, recent SCGF calculations have been performed for 
temperatures above the
critical temperature for a possible phase transition to pairing 
condensation~\cite{rd,frmu:03}. 

In Sec.~II of this work we will review some basic features of the 
SCGF method for the normal Green's function at temperatures above the critical
one for the pairing instability. 
The effects of the anomalous Green's functions will be considered in Sec>~III. 
This leads to the generalized gap equation that corresponds to the homogeneous 
scattering equation for dressed nucleons. 
The analogous problem occurs for two nucleons in the vacuum,
where the solution of the homogeneous scattering equation yields the 
description of the bound two-nucleon state, the deuteron. 
In the nuclear medium, however,
one has to consider a scattering equation with dressed sp propagators. 
If one considers the HF or BHF approximation for the normal
sp propagator, in which the spectral function is replaced by a
$\delta$-function, this gap equation reduces to the usual BCS approximation. 

Results of SCGF calculations above the critical temperature will be discussed
in Sec.~IV.  We will pay special attention to the spectral functions associated
with sp strength around the Fermi energy and discuss, to which extend these 
exhibit typical precursor phenomena for a phase transition to a pairing
condensate~\cite{schnell:99}. In Sect.~IV  we will also investigate how the
distribution of the sp strength modifies the solution of the
gap equation and the subsequent predictions for a phase transition to a
superfluid state of nuclear matter. 
Such investigations have been performed before by Bozek~\cite{boz:99,boz:03} 
using simplified models for the NN interaction. 
These studies suggest an important sensitivity to these
correlations, accompanied by a substantial suppression of the strength of
pairing. A similar conclusion was reached for a realistic interaction in 
Ref.~\cite{dick:99} by studying the phase shifts of dressed nucleons in the
medium that signal the presence of bound pair states as in the case of two free
particles. The inclusion of dressing effects in the study of pairing has also
been  studied in Refs.~\cite{bagr:00,lsz:01,blsz:02} based on the hole-line
expansion of the nucleon self-energy. While also in this work a substantial
reduction of the strength of pairing is observed, the implementation of the
scheme to solve the gap equation relies on approximations that do not conserve
particle number, since they involve the introduction of quasiparticle strength
factors to represent the effect of dressing.
Final conclusions are drawn in Sec.~V.

\section{GREEN'S FUNCTIONS AND T-MATRIX APPROXIMATION}
  
One of the key quantities within the SCGF approach is the sp
Green's function, which can be defined in a grand-canonical formulation for 
both real and imaginary times $t$, 
$t^{\prime}$~\cite{kadanoff}:
\begin{equation}
\label{def_g}
{\mathrm{i}}G({\mathbf{x}},t;{\mathbf{x}}^{\prime},t^{\prime})=
\frac{{\mathrm{tr}}\{\exp[-\beta(H-\mu N)]
{\cal{T}}
[\psi({\mathbf{x}}t) \psi^{\dagger}({\mathbf{x}}^{\prime}t^{\prime})]\}}
{{\mathrm{tr}}\{\exp[-\beta(H-\mu N)]\}}\,,
\end{equation}
where $\cal{T}$ is the time ordering operator, $\beta$ the inverse
temperature, and $\mu$ the chemical potential of the system.
Due to the invariance of the trace under cyclic permutations the one-particle 
Green's function obeys the following quasi-periodicity condition
\begin{equation}
\label{kms}
G({\mathbf{x}},t=0;{\mathbf{x}}^{\prime},t^{\prime})=
-e^{\beta\mu}G({\mathbf{x}},t=-{\mathrm{i}}\beta;{\mathbf{x}}^{\prime},
t^{\prime}).
\end{equation}
For a system invariant under translation in space and time, the propagator
depends only on the differences $|{{\mathbf{x}}-{\mathbf{x}}^{\prime}}|$ and
$t-t^{\prime}$, allowing a Fourier transformation into momentum and energy
variables $k$ and $\omega$. Due to the quasi-periodicity in Eq.~(\ref{kms}), 
the Green's function can be expressed in terms of the Fourier coefficients
$G(k,z_{\nu})$, where 
$z_{\nu}=\frac{\pi\nu}{-{\mathrm{i}}\beta}$ are the (fermion) Matsubara
frequencies with odd integers $\nu$. Since these are related to the spectral 
function $A(k,\omega)$ by
\begin{equation}
\label{spec_g}
G(k,z_{\nu})=\int_{-\infty}^{+\infty}
\frac{{\mathrm{d}}\omega}{2\pi}\, \frac{A(k,\omega)}{z_{\nu}-\omega}\,,
\end{equation}
$G(k,z_{\nu})$ can be continued analytically to all non-real $z$. On the other
hand, the spectral function is related to the imaginary part of the retarded
propagator $G(k,\omega+{\mathrm{i}}\eta)$ by
\begin{equation}
\label{spec_g2}
A(k,\omega)=-2\,{\mathrm{Im}}\,G(k,\omega+{\mathrm{i}}\eta)\,.
\end{equation}
In the limit of the mean-field or quasi-particle approximation the spectral
function is represented by a $\delta$-function and takes the simple form
\begin{equation}
A(k,\omega)=2\pi\delta(\omega + \mu -\varepsilon_k) =2\pi\delta(\omega - \chi_k)
\,,\label{eq:specqp}
\end{equation}
with the quasi-particle energy $\varepsilon_k$ for a particle with momentum $k$
and $\chi_k= \varepsilon_k - \mu$. Note, that for convenience we define the
energy variable relative to the chemical potential $\mu$.

The sp Green's function is obtained as a solution of the Dyson
equation, 
which, for a translationally invariant system, is a simple algebraic
equation
\begin{equation}
\left[\omega +\mu -\frac{k^2}{2m}-\Sigma(k,\omega)\right] 
G(k,\omega) = 1\,,\label{eq:dyson}
\end{equation}
where $\Sigma(k,\omega)$ denotes the complex self-energy. By expanding the self-energy
in terms of one-particle Green's functions it can be demonstrated 
that it inherits all analytic properties of $G$. It it thus possible to write
\begin{equation}
\label{spec_Sigma}
\Sigma(k,\omega)=\Sigma^{HF}(k)-\frac{1}{\pi}\int_{-\infty}^{+\infty}
{\mathrm{d}}\omega^{\prime} \, \frac{{\mathrm{Im}}\Sigma(k,\omega^{\prime}+
{\mathrm{i}}\eta)}
{\omega-\omega^{\prime}}.
\end{equation} 
The next step is to obtain the self energy in terms of the in-medium
two-body scattering $T$ matrix. It is possible to express
${\mathrm{Im}}\Sigma(k,\omega+{\mathrm{i}}\eta)$ in terms of the
retarded $T$ matrix~\cite{kadanoff,frmu:03}
\begin{eqnarray}
\label{im_sigma}
{\mathrm{Im}}\Sigma(k,\omega+{\mathrm{i}}\eta)&=&
\frac{1}{2}\int \frac{{\mathrm{d}}^3k^{\prime}}{(2\pi)^3}
\int_{-\infty}^{+\infty} \frac{{\mathrm{d}}\omega^{\prime}}{2\pi}
\left<{\mathbf{kk}}^{\prime}|
{\mathrm{Im}}T(\omega+\omega^{\prime}+{\mathrm{i}}\eta)|{\mathbf{kk}}^{\prime}\right>
\nonumber \\ && \qquad \times
[f(\omega^{\prime})+b(\omega+\omega^{\prime})]
A(k^{\prime},\omega^{\prime}).
\end{eqnarray}
Here and in the following 
\begin{eqnarray}
f(\omega) & = & \frac{1}{e^{\beta\omega}+1}\,, \nonumber\\
b(\Omega ) & = & \frac{1}{e^{\beta\Omega}-1}\,,\label{eq:fembos}
\end{eqnarray} 
denote the Fermi and Bose distribution functions, respectively. 
The pole in the Bose function $b(\Omega)$ at $\Omega=0$ is 
compensated by a corresponding zero in the $T$ matrix~\cite{alm1,alm95} 
such that the
integrand remains finite as long as the $T$ matrix does not acquire a
pole at this energy. 
Such a pole may occur below a critical temperature $T_C$, a phenomenon
that is often referred to as a pairing instability. We will come back to this
problem below.

The scattering matrix $T$ is to be determined as a solution of the integral
equation
\begin{eqnarray}
\left<{\mathbf{kk}}^{\prime}|T(\Omega+{\mathrm{i}}\eta)|
{\mathbf{pp}}^{\prime}\right> & = &\left<{\mathbf{kk}}^{\prime}|V|
{\mathbf{pp}}^{\prime}\right> + \int
\frac{d^3q\,d^3q^\prime}{\left(2\pi\right)^6} \left<{\mathbf{kk}}^{\prime}|V|
{\mathbf{qq}}^{\prime}\right>G^0_{\mathrm{II}}(\mathbf{qq}^\prime,\Omega+i\eta)
\nonumber \\ &&\quad\quad\times
\left<{\mathbf{qq}}^{\prime}|T(\Omega+{\mathrm{i}}\eta)|
{\mathbf{pp}}^{\prime}\right>\,,\label{eq:tscat0}
\end{eqnarray}
where
\begin{equation}
\label{two_pp}
G^0_{\mathrm{II}}(k_1,k_2,\Omega+i\eta)=
\int_{-\infty}^{+\infty}\frac{{\mathrm{d}}\omega}{2\pi}
\int_{-\infty}^{+\infty}\frac{{\mathrm{d}}\omega^{\prime}}{2\pi}
A(k_1,\omega)A(k_2,\omega^{\prime})
\frac{1-f(\omega)-f(\omega^{\prime})}
{\Omega-\omega-\omega^{\prime}+i\eta}\,.
\end{equation}
stands for the two-particle Green's function of two non-interacting but
dressed nucleons.

The Green's function method yields a hierarchy of relations for the 
$N$-particle Green's functions.
The Dyson equation for the one-particle Green's
function involves the two-body potential as well as the 
two-particle Green's function
$G_{\mathrm{II}}({\mathbf{x}},t;\cdots;{\mathbf{x}}^{\prime\prime\prime},
t^{\prime\prime\prime})$.
In general, the equation of motion for the $N$-particle propagator will
be coupled to the $(N+1)$-particle propagator, if the Hamiltonian
contains a two-body interaction. In the self-consistent $T$-matrix approach,
as outlined above, one ignores the effects of $N$-particle Green's functions
with $N$ larger equal to three, but solves the coupled equations for the one-
and two-body Green's functions in a self-consistent way.

In order to allow for an efficient solution of the two-body scattering equation,
one expresses the two-particle Green's function in (\ref{two_pp}) as a function 
of the total momentum   
${\mathbf{P}}=\frac{1}{2}({\mathbf{k_1}}+{\mathbf{k_2}})$ and the relative
momentum ${\mathbf{q}}=\frac{1}{2}({\mathbf{k_1}}-{\mathbf{k_2}})$. Using the
usual angle-average approximation for the angle between ${\mathbf{P}}$ and 
${\mathbf{q}}$ (see \textit{e.g.}~\cite{angleav} 
for the accuracy of this approximation), 
the two-particle Green's function can be written as a function 
of the length of these two-vectors, $P$ and $q$, only. This approximation leads
to a decoupling of partial waves with different total angular momentum $J$.  
Therefore the integral equation (\ref{eq:tscat0}) reduces to an integral
equation in only one dimension of the form
\begin{eqnarray}
\left<q|T^{JST}_{ll^{\prime}}(P,\Omega+{\mathrm{i}}\eta)|q^{\prime}\right>& = &
\left<q|V^{JST}_{ll^{\prime}}|q^{\prime}\right>
+\sum_{l^{\prime \prime}} \frac{2}{\pi}\int_0^{\infty} 
{\mathrm{d}} k^{\prime}\,k^{\prime 2}
\left<q|V^{JST}_{ll^{\prime \prime}}|k^{\prime}\right>
G^0_{\mathrm{II}}(P,\Omega+{\mathrm{i}}\eta,k^{\prime})
\nonumber\\
&&\quad \quad \times
  \left<k^{\prime}|T^{JST}_{l^{\prime \prime}l^{\prime}}
(P,\Omega+{\mathrm{i}}\eta)|q^{\prime}\right>.
\label{T_waves}
\end{eqnarray}
The summation of the partial waves,
\begin{equation}
\left<{\mathbf{kk}}^{\prime}\right|
	{\mathrm{Im}}\,T(\Omega+{\mathrm{i}}\eta)
	\left|{\mathbf{kk}}^{\prime}\right>
	=
	\frac{1}{4\pi}\sum_{(JST)l}(2J+1)(2T+1)
	\left<q({\mathbf{k}},{\mathbf{k}}^{\prime})\right|{\mathrm{Im}}\,
	T^{JST}_{ll}(P({\mathbf{k}},{\mathbf{k}}^{\prime}),\Omega+{\mathrm{i}}\eta)
	\left|q({\mathbf{k}},{\mathbf{k}}^{\prime})\right>,
\label{sum_parwaves}
\end{equation}
yields the $T$ matrix in the form that is needed in
Eq.~(\ref{im_sigma}).
Finally, the Hartree-Fock contribution has to be added to the real
part of $\Sigma$
\begin{equation}
\label{hf_sigma}
\Sigma^{HF}(k)
=
\frac{1}{8\pi}\sum_{(JST)l}(2J+1)(2T+1)
\int \frac{{\mathrm{d}}^3k^{\prime}}{(2\pi)^3}
\left<q({\mathbf{k}},{\mathbf{k}}^{\prime})\right|
V^{JST}_{ll}
\left|q({\mathbf{k}},{\mathbf{k}}^{\prime})\right>
n(k^{\prime}),
\end{equation}
where $n(k)$ is the correlated momentum distribution
\begin{equation}
\label{occupation}
n(k)=
\int_{-\infty}^{+\infty} \frac{{\mathrm{d}}\omega}{2\pi}
f(\omega)
A(k,\omega).
\end{equation}
Note, that Eq.~(\ref{hf_sigma}) corresponds to a 
generalized Hartree-Fock contribution, since the
full one-particle spectral function is employed.

\section{PAIRING IN THE T-MATRIX APPROXIMATION}
For temperatures below the critical temperature for a transition to a
superfluid one has to
supplement the evaluation of the normal Green's function $G(k,\omega)$ with the
anomalous Green's function $F(k,\omega)$.  
While the self-consistent inclusion of ladder diagrams has reached quite a 
sophistication, it remains to fully account for the possibility of a pairing
solution in such calculations for realistic NN interactions.
In this paper we present a first step towards such a complete scheme by
including the full self-consistent dressing due to normal self-energy terms
generated by ladder diagrams, in the calculation of the anomalous self-energy
and the solution of the corresponding generalized gap equation.

The inclusion of the anomalous Green's function $F(k,\omega)$ yields a
modification of the normal Green's function in the superfluid phase
that can be written as~\cite{migdal,schrieffer,mahan,diva:05}
\begin{eqnarray}
G_s(\mathbf{k},\omega+i\eta)& = & G(\mathbf{k},\omega+i\eta) -  
G(\mathbf{k},\omega+i\eta) \Delta(\mathbf{k}) F(\mathbf{k},\omega +
i\eta)\nonumber \\
F(\mathbf{k},\omega + i\eta) & = & G(-\mathbf{k},-\omega -i\eta)
G_s(\mathbf{k},\omega+i\eta) \Delta(\mathbf{k})\,.\label{eq:coupdys}
\end{eqnarray}
under the assumption that the anomalous part of self-energy $\Delta$ does not 
depend on the energy. 
Therefore the full Green's function can be obtained as
\begin{equation}
G_s(\mathbf{k},\omega+i\eta) = \frac{1}{G(\mathbf{k},\omega+i\eta)^{-1} +
\Delta^2(\mathbf{k})G(-\mathbf{k},-\omega -i\eta)}\,.
\end{equation}
These equations must be supplemented with the definition of the anomalous
self-energy
\begin{equation}
\Delta(\mathbf{p}) = \int \frac{d\omega}{2\pi} 
\int \frac{d^3k}{\left(2\pi\right)^3} 
\left<\mathbf{p}|V|\mathbf{k}\right>
2 {\mathrm{Im}}F(\mathbf{k},\omega + i\eta)f(\omega)\,.
\end{equation}
If one employs Eq.~(\ref{eq:coupdys}) and uses the representation of the Green's
function $G$ in terms of the spectral function in Eqs.~(\ref{spec_g}) and
(\ref{spec_g2}), 
supplemented by a corresponding definition of a spectral function
for the total Green's function
\begin{equation}
\label{spec_g2s}
A_s(k,\omega)=-2\,{\mathrm{Im}}\,G_s(k,\omega+{\mathrm{i}}\eta)\,,
\end{equation} 
the expression for the self-energy $\Delta$ can be rewritten~\cite{boz:99}
in a partial wave expansion
\begin{equation}
\Delta^{JST}_l(p) = \sum_{l^{\prime}} \frac{2}{\pi}\int_0^{\infty} 
{\mathrm{d}} k\,k^{2}
\int_{-\infty}^{+\infty}\frac{{\mathrm{d}}\omega}{2\pi}
\int_{-\infty}^{+\infty}\frac{{\mathrm{d}}\omega^{\prime}}{2\pi}
\left<p|V^{JST}_{ll^{\prime}}|k\right>
A(k,\omega)A_s(k,\omega^{\prime})
\frac{1-f(\omega)-f(\omega^{\prime})}
{-\omega-\omega^{\prime}}\Delta^{JST}_{l^\prime}(k)\,.
\label{eq:gappw}
\end{equation}
If we ignore for a moment the difference between the spectral functions $A$ and
$A_s$, we see that this equation for the self-energy $\Delta$ corresponds to 
the homogeneous scattering equation for the $T$-matrix in (\ref{eq:tscat0}) 
at energy $\Omega=0$ and center-of-mass momentum $P=0$. 
This means that a non-trivial solution of Eq.~(\ref{eq:gappw}) is
obtained if and
only if the scattering matrix $T$ generates a pole at energy $\Omega=0$, which 
reflects a bound two-particle state. This is precisely the condition for the
pairing instability discussed above, demonstrating that this treatment of 
pairing correlations is compatible with the $T$-matrix approximation in the
non-superfluid regime discussed in Sec.~II. 

We may also consider Eq.~(\ref{eq:gappw}) in the limit in which we approximate
the spectral functions $A(k,\omega)$ and $A_s(k,\omega)$ by the corresponding
mean-field and BCS approximation. The expression for the normal spectral
function has been presented already in Eq.~(\ref{eq:specqp}). The BCS
approximation for the spectral function yields
\begin{equation}
A_s(k,\omega) = 2\pi\left(\frac{E_k+\chi_k}{2E_k}\delta(\omega-E_k) +
\frac{E_k-\chi_k}{2E_k}\delta(\omega+E_k)\right),\label{eq:specbcs}
\end{equation}
with the quasi-particle energy
\begin{equation}
E_k = \sqrt{\chi_k^2 + \Delta^2(k)}\,.\label{eq:eqp}
\end{equation} 
Inserting these approximations for the spectral function into
Eq.~(\ref{eq:gappw}) and taking the limit $T=0$ reduces to the usual
BCS gap equation
\begin{equation}
\Delta^{JST}_l(p) = \sum_{l^{\prime}} \frac{2}{\pi}\int_0^{\infty} 
{\mathrm{d}} k\,k^{2} \left<p|V^{JST}_{ll^{\prime}}|k\right>\frac{1}{-2E_k}
\Delta^{JST}_{l^\prime}(k)\,.
\label{eq:gapbcs}
\end{equation}
Therefore we can consider Eq.~(\ref{eq:gappw}) as a generalization of the usual
gap equation. It accounts for the spreading of sp strength
leading to a generalization of the form
\begin{equation}
\frac{1}{-2E_k}\quad \to \quad  
\int_{-\infty}^{+\infty}\frac{{\mathrm{d}}\omega}{2\pi}
\int_{-\infty}^{+\infty}\frac{{\mathrm{d}}\omega^{\prime}}{2\pi} 
A(k,\omega)A_s(k,\omega^{\prime})
\frac{1-f(\omega)-f(\omega^{\prime})}
{-\omega-\omega^{\prime}}\,.\label{def:zk0}
\end{equation}

\begin{figure}[bth]
\begin{center}
\includegraphics[origin=bl,width=8.0cm]{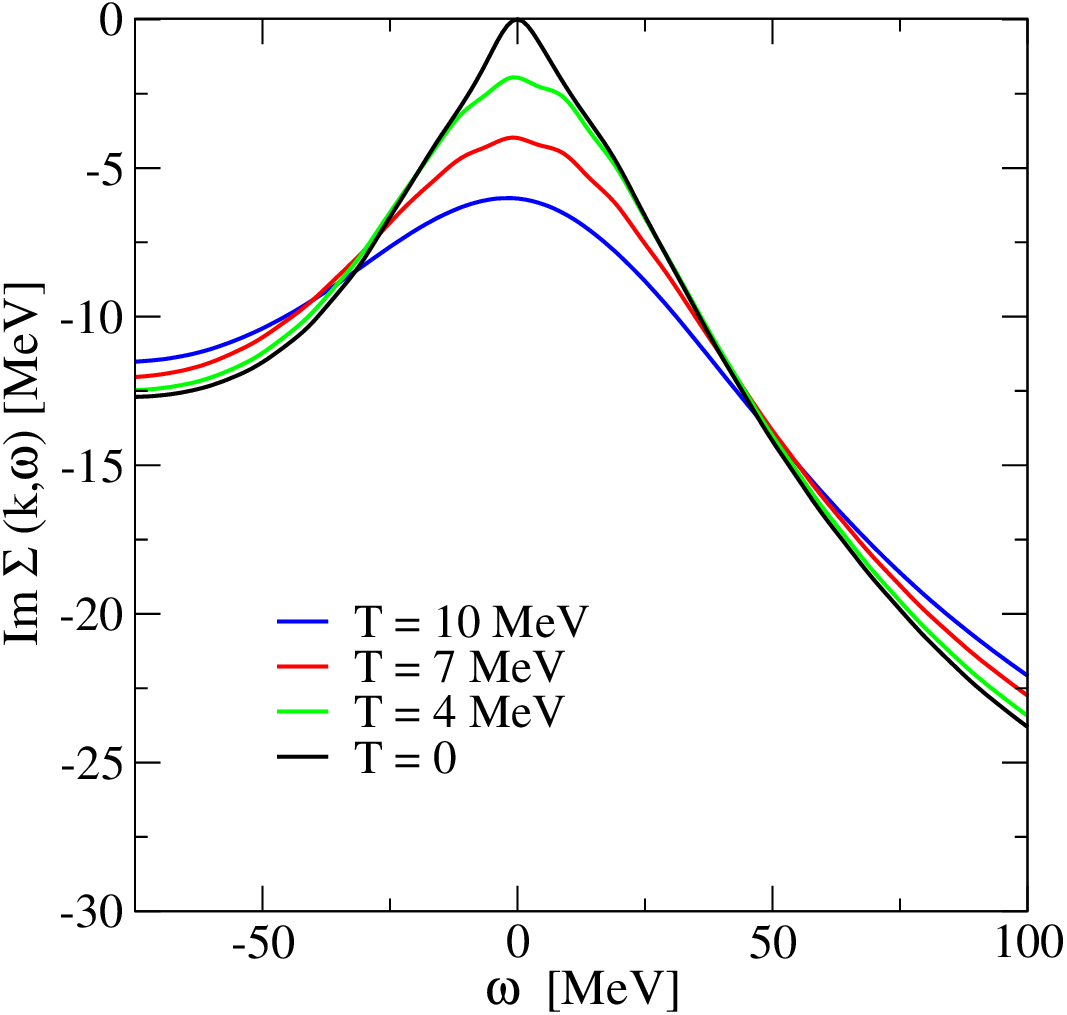}
\caption{Imaginary part of the retarded self-energy for nucleons with momentum
$k$ = 225 MeV/c in symmetric nuclear matter at the empirical saturation density
($\rho$ = 0.16 fm$^{-3}$). Results for temperatures $T$ larger than or equal 
to 4 MeV are directly determined by SCGF calculations, 
while the $T=0$ result originates from an
extrapolation. All results were obtained using the CDBonn interaction. 
\label{fig:imagself}}
\end{center}
\end{figure}

\section{RESULTS AND DISCUSSION}
\subsection{SCGF above the critical temperature}
In the first part of this section we will focus the attention to the discussion
of SCGF calculations for symmetric nuclear matter and pure neutron matter at
temperatures above the critical temperature $T_c$ for a phase 
transition to a pairing condensate. 
This means that we solve Eqs.~(\ref{T_waves})
- (\ref{occupation}) in an iterative procedure until a self-consistent solution
is obtained. Some details of this procedure have been published in
\cite{frmu:03} and \cite{thesis:fr}. 

As a typical example we present in Fig.~\ref{fig:imagself} results for the
imaginary part of the retarded self-energy $\Sigma (k,\omega)$ for nucleons
with a fixed momentum $k$ as a function of the energy variable $\omega$. All
the results displayed in this figure have been determined for symmetric nuclear
matter at the empirical saturation density using the CDBonn~\cite{cdbonn}
interaction. The energy scale in this figure has been constrained to energies
around the Fermi energy ($\omega = 0$), since the self-energy is most 
sensitive to the temperature for these energies. 
The results for temperatures larger than or equal to
4 MeV, which are all above $T_c$ (see below), have been obtained directly 
from SCGF calculations. They exhibit a rather smooth dependence on the
temperature, so that an extrapolation to temperatures below $T_c$ appears
feasible. 
As an example of such an extrapolation we show the $T=0$ result. This
extrapolation has been done with the constraint that the imaginary part of the
self-energy for $T=0$ vanishes at $\omega$ =0. 

\begin{figure}[bth]
\begin{center}
\includegraphics[origin=bl,width=8.0cm]{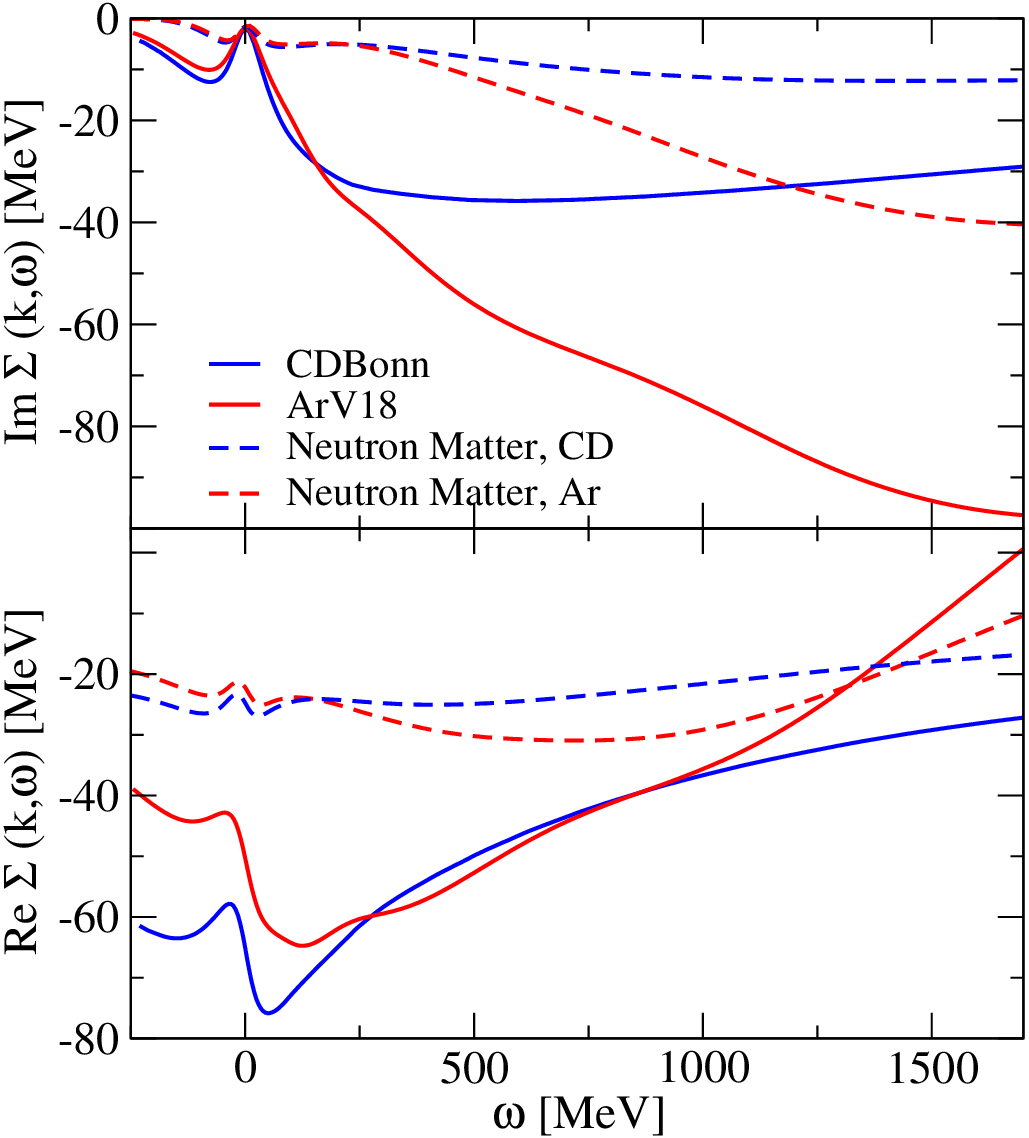}
\caption{Imaginary (upper panel) and real part (lower panel) of the retarded 
self-energy for nucleons with momentum
$k$ = 225 MeV/c in symmetric nuclear matter at the empirical saturation density
($\rho$ = 0.16 fm$^{-3}$). Results obtained for CDBonn interaction are compared
to those resulting from SCGF calculations using ArV18. 
Also included are results
for neutrons with the same momentum in neutron matter at $\rho$ = 0.08 fm$^{-3}$. 
The temperature in all these calculations was fixed at $T$=5 MeV.
\label{fig:compself}}
\end{center}
\end{figure}

The imaginary part of the nucleon self-energy is also displayed in the upper
panel of Fig.~\ref{fig:compself}. The purpose of this figure is to visualize
some differences between various models of the NN interaction and between
symmetric nuclear matter and pure neutron matter. 
Therefore we consider a larger
interval for the energy variable $\omega$. The imaginary parts of the
self-energy derived from the CDBonn interaction and the Argonne V18 (ArV18)
interaction~\cite{arv18} are very similar at energies around $\omega=0$. At
those energies the ArV18 yields a slightly weaker imaginary part than CDBonn.
The differences get larger at positive values of $\omega$, where the imaginary
part of the self-energy derived from ArV18 reaches a minimum of around -100 MeV
at an energy $\omega$ around 1.7 GeV. The minimum for the CDBonn interaction
is only about -35 MeV and occurs at energies $\omega$ around 0.5 GeV. 

This is another indication of the feature, that a fit of a local interaction,
like ArV18, to NN phase shifts yields a larger amount of NN correlations than a
fit of a non-local relativistic meson exchange model like the
CDBonn fitting identical phase shifts~\cite{mupo:00,gad:01}. In short: The 
ArV18 is a stiffer interaction than CDBonn. 
A further illustration is provided by a Hartree-Fock
calculation for nuclear matter at the empirical saturation density 
that yields a total energy 30 MeV per nucleon for the ArV18, while CDBonn 
generates 5 MeV per nucleon~\cite{mupo:00}.
This implies that the generalized Hartree-Fock contribution to the
self-energy in Eq.~(\ref{spec_Sigma}), defined in Eq.~(\ref{spec_Sigma}),
is more repulsive for ArV18, and a larger part of the attraction is
provided by the energy-dependent contribution to the real part of the
self-energy. This is immediately obvious, since 
the energy-dependent contribution to the real
part of $\Sigma$ is connected to the imaginary part by a dispersion relation.
The lower panel of Fig.~\ref{fig:compself}, which
displays results for the real part of the self-energy, illustrates this 
observation: The energy dependence is
larger for ArV18 as compared to CDBonn. 
The weaker attraction of the self-energy
derived from CDBonn (for most values of $\omega$) reflects the less repulsive
contribution of the generalized Hartree-Fock contribution.

Figure~\ref{fig:compself} also displays results for the real and 
imaginary part of
the self-energy for neutrons with the same momentum ($k$ = 225 MeV) in pure
neutron matter. The density of neutron matter considered in this figure is one
half of the empirical saturation density of nuclear matter, which implies that
these systems have the same Fermi momentum. The imaginary part of the
self-energy in neutron matter is weaker than the corresponding one for
symmetric nuclear matter, reflecting the dominance of proton-neutron
correlations. For both interactions a minimum is obtained around
1.7 GeV. At these high energies the absolute value for the imaginary part of 
the self-energy is about a factor three larger for ArV18 than for CDBonn.
This means that the distribution of sp strength to high energies
due to central short-range correlations is much stronger for the local ArV18
interaction than for the non-local meson-exchange model CDBonn. The results for
the lower energies are closer to each other.
The differences in the amount of NN correlations is also reflected in the
occupation probability $n(k)$ defined in (\ref{occupation}). For symmetric
nuclear matter at saturation density we obtain for $n(k=0)$ the values 0.89 and
0.87 for CDBonn and ArV18, respectively. The corresponding value for neutron
matter are $n(k=0)$ = 0.968 and 0.963. 
Earlier non-self-consistent calculations with older NN interactions tended to 
yield values of 0.83 for this quantity~\cite{vond1} in nuclear matter, 
with self-consistency raising the number to 0.85~\cite{rd}.

\begin{figure}[bth]
\begin{center}
\includegraphics[origin=bl,width=7.0cm]{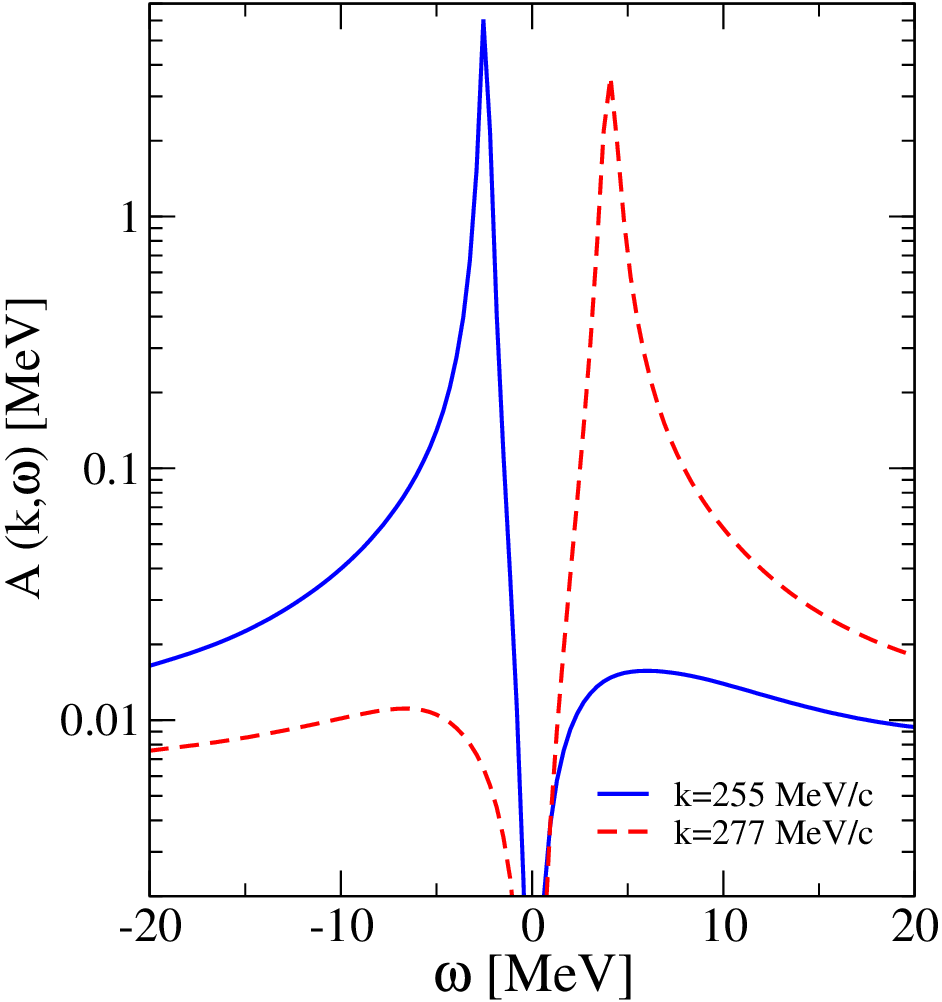}
\caption{Spectral function for nucleons in symmetric nuclear matter at the
empirical saturation density. Results of SCGF calculations have been
extrapolated to $T=0$. The CDBonn interaction has been used.
\label{fig:specfunc}}
\end{center}
\end{figure}

Examples for spectral functions $A(k,\omega)$ are displayed in
Fig.~\ref{fig:specfunc}. Again we consider the case of symmetric nuclear matter
at the empirical saturation density and use the CDBonn interaction. As examples
we consider two momenta $k=255$ MeV/c and $k=277$ MeV/c, which are below and
above the Fermi-momentum $k_F$, respectively. For the momentum $k < k_F$ one
finds the dominant peak at an energy below the Fermi energy $\omega = 0$ and a
much smaller maximum at $\omega$ larger than zero. 
For momenta $k>k_F$ the dominant
quasi-particle peak is located at positive values of $\omega$ and a second
maximum occurs at $\omega < 0$. 
Since this feature of two maxima in the spectral
function is reminiscent of the two poles that 
are present in the BCS approximation to the Green's function, 
it has been discussed as the formation of a pseudo-gap or as a
precursor phenomenon to a pairing condensate~\cite{boz:99}. 
We note, however, that
our calculations only exhibit this feature of two pronounced maxima in the 
spectral function, if we consider rather low temperatures,
in particular $T<T_c$. The examples
displayed in Fig.~\ref{fig:specfunc} originate from an extrapolation of the
self-energy to $T=0$. This is different from results obtained with simplified
interactions, as they are used \textit{e.g.} 
in Ref.~\cite{boz:99}. The interaction
employed by Bo\.zek yields an imaginary part of the self-energy, which is 
different from zero in a much smaller energy interval than the realistic
calculations considered here. Due to this difference spectral functions with
two maxima are obtained also at temperatures above $T_c$ for this model
interaction. 

\subsection{Pairing correlations}
As a first step towards the study of pairing correlations, we consider
the usual BCS approach. This means that we solve the gap equation
(\ref{eq:gapbcs}) assuming a spectrum of sp energies
$\varepsilon(k)=\chi_k + \mu$, which we determine from the quasi-particle 
energies
\begin{equation}
\varepsilon(k) = \frac{k^2}{2m} +  \mathrm{Re} \Sigma(k,\varepsilon(k)-\mu)\,.
\label{eq:epsqp}
\end{equation}
In this equation $\mathrm{Re} \Sigma$ denotes the result of a SCGF calculation
extrapolated to $T=0$. Such spectra of quasi-particle energies are rather
similar to the sp spectra used in other work. 
The corresponding BCS calculations therefore involve
the usual procedure as it has been applied \textit{e.g.} in
Refs.~\cite{dehj:03,ars:90,vgdpr:91,bbl:92,tt:93,bls:95}. 

\begin{figure}[bth]
\begin{center}
\includegraphics[origin=bl,width=9.0cm]{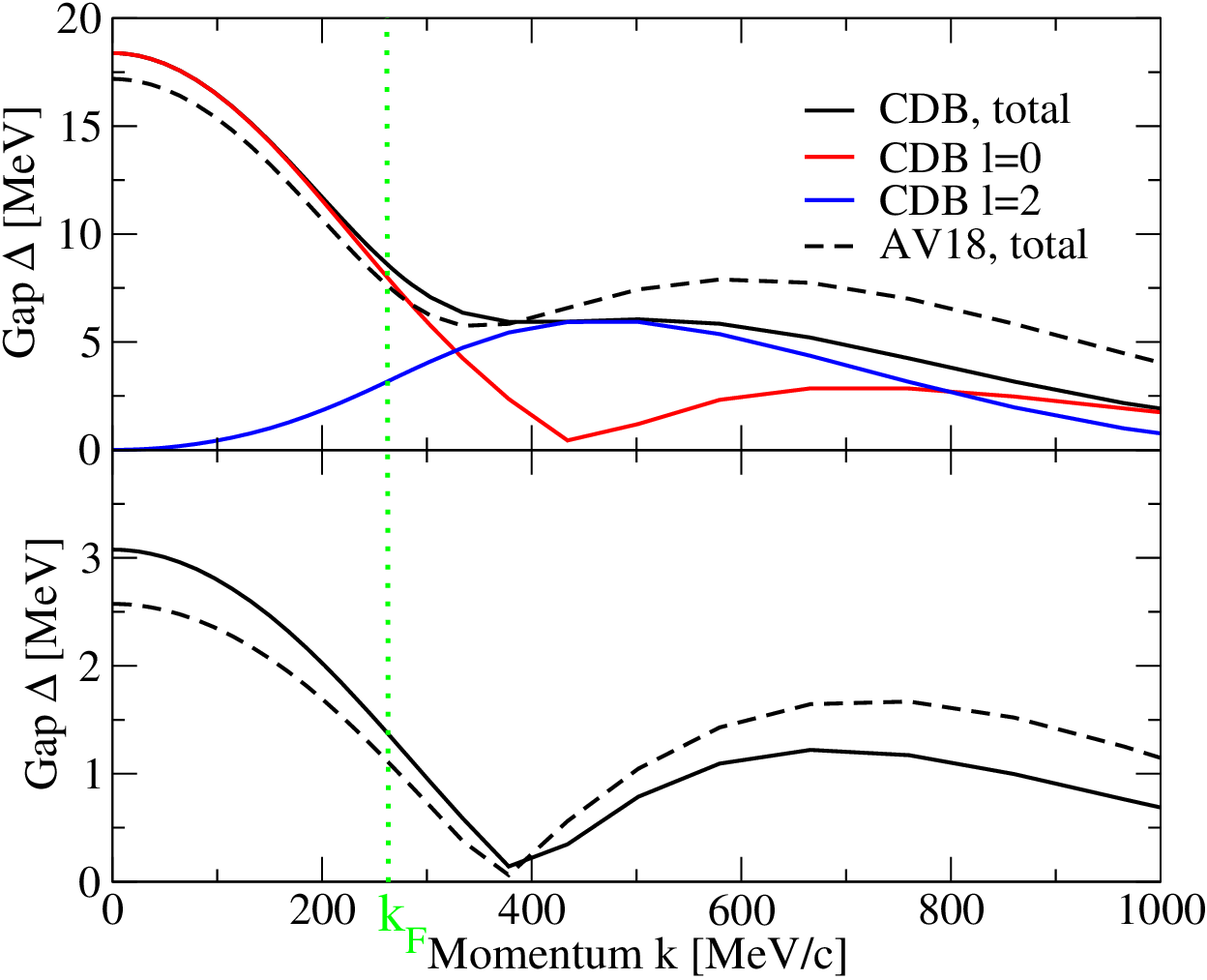}
\caption{Results for the gap-functions $\vert \Delta_l(k)\vert$ for symmetric
nuclear matter ($\rho$ = 0.16 fm$^{-3}$, upper panel) and pure neutron matter
($\rho$ = 0.08 fm$^{-3}$, lower panel) obtained from a solution of the BCS
equation (\protect{\ref{eq:gapbcs}}) using the CDBonn and ArV18 interactions 
at $T=0$. The dotted line identifies the Fermi-momentum $k_F$.
\label{fig:gap3s2}}
\end{center}
\end{figure}

Results for the gap-functions $\vert \Delta_l(k)\vert$ are displayed in  
Fig.~\ref{fig:gap3s2}. The upper panel of this figure shows results for
symmetric nuclear matter at saturation density. 
The partial wave that yields
the largest value for $\Delta$ and is therefore the relevant one in
this case, is the $^3S_1-^3D_1$ channel describing the proton-neutron 
interaction. 
We therefore display the absolute
values of the gap-functions for $l=0$ and $l=2$ ($\Delta_0$ and $\Delta_2$) as
well as the total gap-function $\Delta = \sqrt{\Delta_0^2+\Delta_2^2}$ as a
function of the momentum $k$. Below we will mainly consider the value of the
gap-function $\Delta$ at the Fermi momentum $k_F$. We also compare in this
figure the results obtained from CDBonn with those from ArV18. For
smaller values of $k$ the CDBonn yields larger values for the gap function,
while ArV18 leads to larger gap values for momenta larger than $k= 400$
MeV/c. This is true for the $l=0$ component as well as the $l=2$ component and
consequently also for the total result. This feature at large values of $k$ is
in line with our observation made above, that ArV18 tends to produce a larger
amount of correlations at high momenta and large energies. For lower momenta,
however, CDBonn yields larger gap-functions. Therefore the gap (at the Fermi
momentum) resulting from a BCS calculation, which uses CDBonn (8.6 MeV) is 
larger than the corresponding value calculated for ArV18 (7.6 MeV), although
ArV18 tends to produce more short-range correlations than CDBonn.

The situation is quite similar for the case of neutron-neutron pairing in pure
neutron matter, which is displayed in the lower part of Fig.~\ref{fig:gap3s2}.
In this case the pairing effects are dominated by the $^1S_0$ partial wave. At
high momenta we obtain larger values for the gap function using ArV18, whereas
CDBonn yields larger values for $\Delta(k)$ at low momenta. Therefore the value
$\Delta(k_F)$ is larger for CDBonn (1.4 MeV) than for ArV18 (1.1 MeV). These
values for the neutron-neutron pairing gap are, however, much lower than the
corresponding values for proton-neutron pairing at the same Fermi momentum. On
one hand this sounds natural, as we know that the proton-neutron interaction is
stronger than the neutron-neutron interaction, leading to a bound deuteron and 
to more correlations (see above). On the other hand, however, one observes
the effects of proton-proton and neutron-neutron pairing in finite nuclei, 
while there is hardly any trace of proton-neutron pairing effects in nuclei.

\begin{figure}[bth]
\begin{center}
\includegraphics[origin=bl,width=8.0cm]{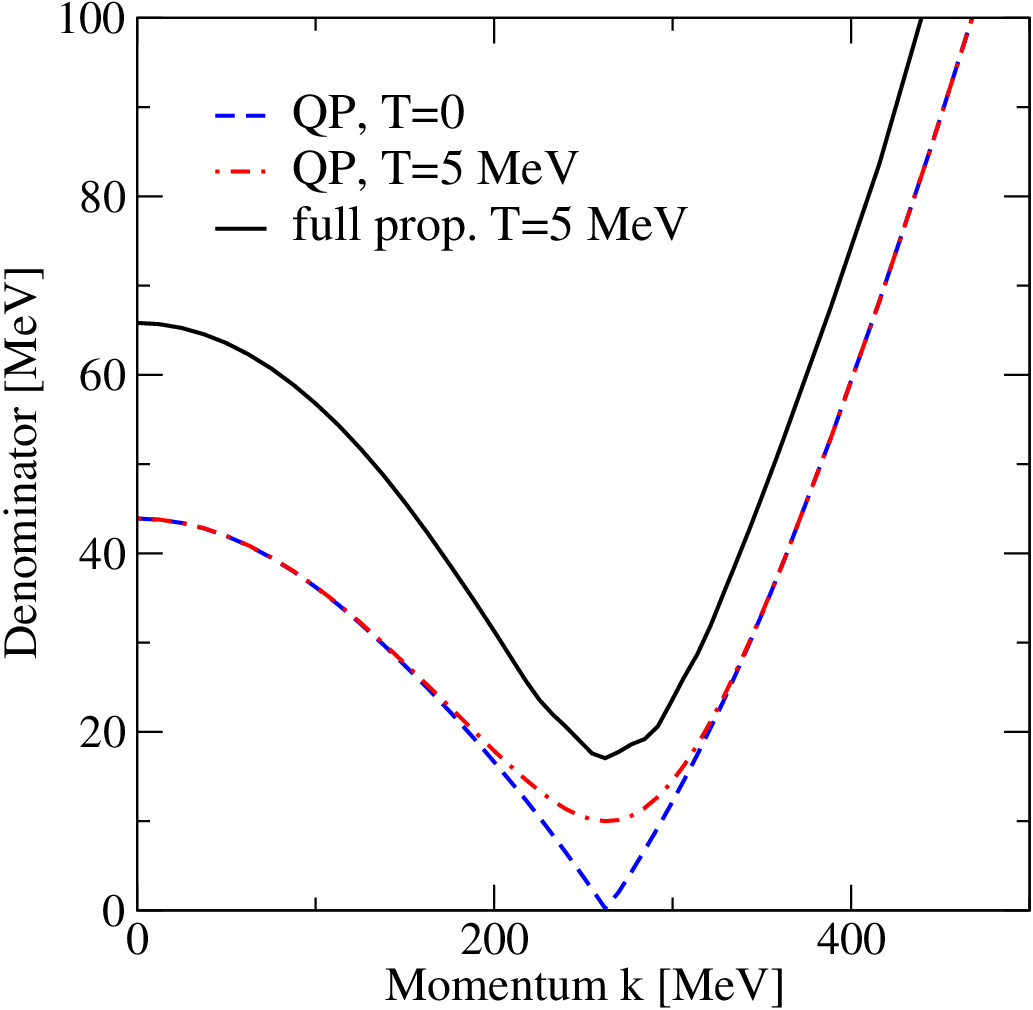}
\caption{The quantity $\widetilde\chi_k$ defined in
Eq.~(\protect{\ref{def:zk1}}), which represents the energy denominator for the
propagator of two nucleons in the medium. Results are displayed for the
quasi-particle approximation in the limit $T=0$ (QP, T=0), the quasi-particle
approximation for finite temperature $T=5$ MeV (QP, T=5 MeV) and the dressed
propagator resulting from SCGF calculations (full prop., T=5 MeV). All results
displayed in this figure were obtained for symmetric nuclear matter at density
$\rho$ = 0.16 fm$^{-3}$, using CDBonn interaction.
\label{fig:denominator}}
\end{center}
\end{figure}

As a next step, we now try to consider the effects of temperature and
short-range correlations in the solution of the gap equation. For that purpose
we will reconsider the two-particle propagator of Eq.~(\ref{def:zk0}) 
but replace
the spectral function of the superfluid phase, $A_s(k,\omega^\prime)$, by the
corresponding one for the normal phase, $A(k,\omega^\prime)$. If we consider
this propagator in the limit of the mean-field approximation 
($A(k,\omega)=\delta(\omega-\varepsilon_k)$) at $T=0$, it
reduces to an energy denominator of the form
\begin{equation}
\frac{1}{-2\widetilde\chi_k} = :
\int_{-\infty}^{+\infty}\frac{{\mathrm{d}}\omega}{2\pi}
\int_{-\infty}^{+\infty}\frac{{\mathrm{d}}\omega^{\prime}}{2\pi} 
A(k,\omega)A(k,\omega^{\prime})
\frac{1-f(\omega)-f(\omega^{\prime})}
{-\omega-\omega^{\prime}} \quad \stackrel{{\rm mf, }T=0}{\longrightarrow} \quad
\frac{1}{-2\vert\chi_k\vert} 
\,.\label{def:zk1}
\end{equation}
This means that the energy $\widetilde\chi_k$ has been defined in this
equation to exhibit the effects of finite temperature and correlations on the
two-particle propagator. Figure~\ref{fig:denominator} displays results for this
quantity $\widetilde\chi_k$, the inverse of this propagator multiplied by -2,
and  compares it with $\vert\chi_k\vert$ using the corresponding quasi-particle
energies. The dashed-dotted line represents the effects of temperature, 
\textit{i.e.}
the propagator has been calculated using the quasi-particle approach for the
spectral function and the Fermi function for the temperature under
consideration, while the solid line accounts for finite temperature and
correlation effects.  
The finite temperature yields an enhancement of the effective sp
energy $\widetilde\chi_k$ for momenta around the Fermi-momentum, only. 
Including in addition
the effects of correlations in the propagator, we obtain larger values for 
$\widetilde\chi_k$ for all momenta. This corresponds to the well known feature
that a finite temperature yields a depletion of the occupation probability of
sp states only for momenta just below the Fermi momentum, while 
strong short-range correlation provide such a depletion for all momenta of the
Fermi sea.

\begin{figure}[bth]
\begin{center}
\includegraphics[origin=bl,width=10.0cm]{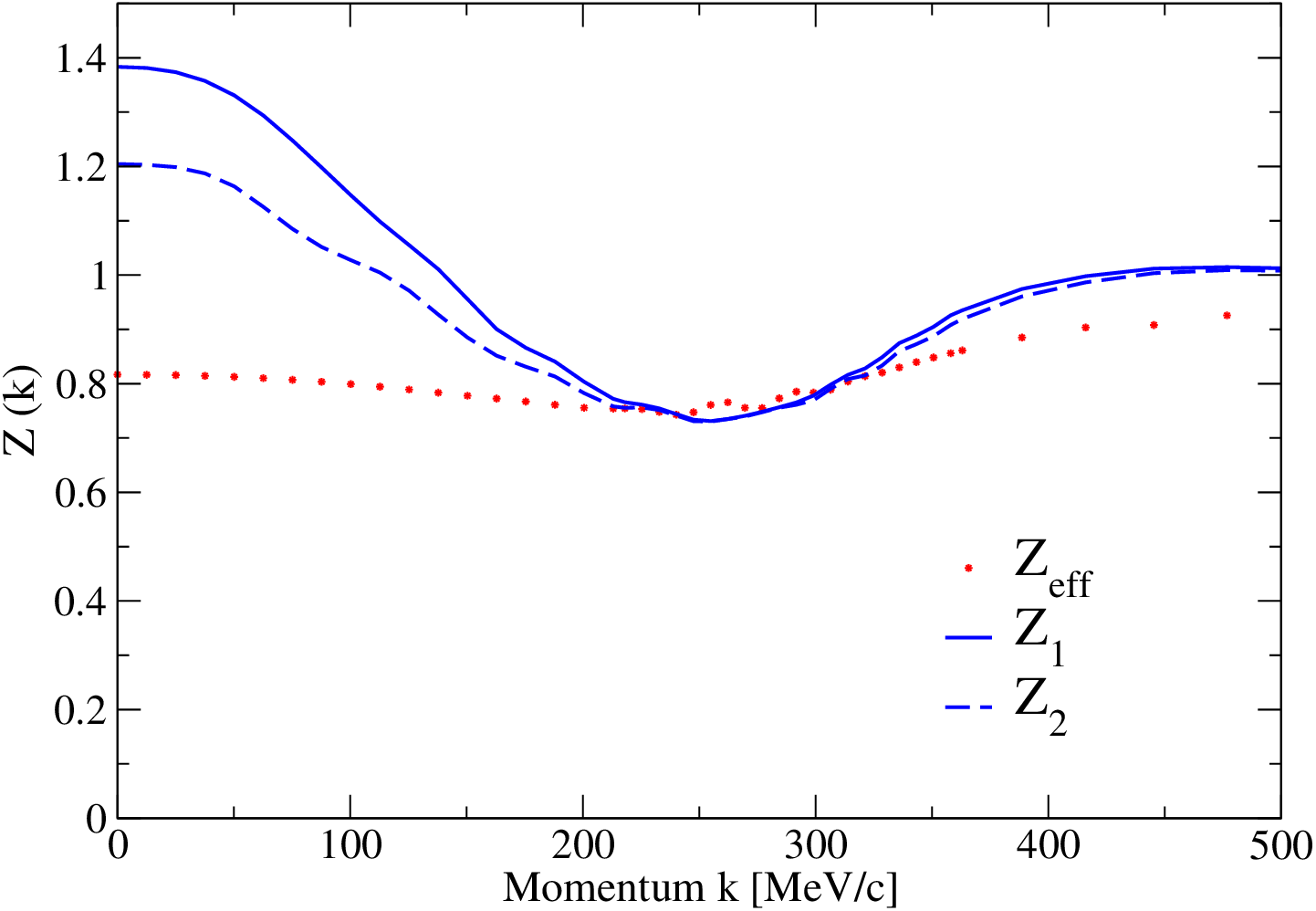}
\caption{Effective renormalization constants for the sp strength
located in the quasi-particle peak as derived from 
Eq.~(\protect{\ref{def:zk2}})
are compared to the estimates of Eq.~(\protect{\ref{eq:zk1}}) and 
Eq.~(\protect{\ref{eq:zk2}}). Further details as in
Fig.~\protect\ref{fig:denominator}.
\label{fig:zfact}}
\end{center}
\end{figure}

Now one may try to describe this depopulation of the sp strength
due to the short-range correlations by a factor $Z$, which represents the part
of the sp strength contained in the quasi-particle peak.
Considering the propagator of Eq.~(\ref{def:zk1}) we define
\begin{equation}
\frac{1}{-2\widetilde\chi_k} (\mbox{T, dressed}) = \frac{Z_{eff}^2(k)}
{-2\chi_k} (\mbox{T, QP})\,,\label{def:zk2}
\end{equation} 
which means that we determine $Z_{eff}$ \textit{e.g.} 
from the ratio of the results
displayed by the solid and the dashed-dotted line in
Fig.~\ref{fig:denominator}. Results for such a strength factor are displayed in
Fig.~\ref{fig:zfact}. Typical values for $Z_{eff}(k)$ are around 0.8 to 0.9 and
show only a weak dependence on $k$.  
This demonstrates that the effects of correlations may
be expressed in terms of a renormalization factor $Z(k)$ to be used in the
usual BCS equation. Such renormalization effects have been discussed by
Bo{\.{z}}ek~\cite{boz:03,boz:00} and Baldo \textit{et al.}~\cite{blsz:02}.  
Note, however, that the renormalization factor $Z_{eff}(k)$ is defined by
Eq.~(\ref{def:zk2}) and the complete distribution of sp strength is
required to calculate it. It will be interesting to examine, to which extend
this factor can be approximated by simpler estimates for the strength located
in the quasi-particle peak.

For that purpose we present in Fig.~\ref{fig:zfact} also results from simple
estimates of this strength distribution. If one assumes that the imaginary part
of the self-energy is a constant, not depending on the energy variable 
$\omega$, one can estimate this renormalization factor by
\begin{equation} 
Z_1(k) = \frac{1}{1-\frac{d\mathrm{Re}\Sigma}{d\omega}}\,.\label{eq:zk1}
\end{equation}
Note, however, that the real-part of the self-energy $\mathrm{Re}\Sigma$
calculated in SCGF yields negative slopes as a function of energy for various
momenta and energies (see Fig.~\ref{fig:compself}). Therefore the values for
$Z_1(k)$ yield values larger than 1 over a wide range of momenta, which is very
different from the corresponding values for $Z_{eff}(k)$. 
One may try to improve
the estimate for the strength factor by accounting for an energy dependence of
the imaginary part of the self-energy $\mathrm{Im}\Sigma$. This leads
to~\cite{jlm} 
\begin{equation} 
Z_2(k) = \frac{1}{\sqrt{\left(1-\frac{d\mathrm{Re}\Sigma}{d\omega}\right)^2 +
\left(\frac{d\mathrm{Im}\Sigma}{d\omega}\right)^2}}\,.\label{eq:zk2}
\end{equation}
This improvement also does not lead to results that are consistent
with $Z_{eff}(k)$ (see Fig.~\ref{fig:zfact}). This means that we expect results
for the generalized gap-equation (\ref{eq:gappw}) which are similar to those
obtained in Refs.~\cite{boz:03,boz:00,blsz:02} using appropriate values for
$Z_{eff}(k)$ but we cannot give a simple reliable scheme to estimate the values
for the renormalization constants. 
We therefore determine the dressed two-particle propagator in 
Eq.~(\ref{def:zk1}) within the SCGF approximation by 
extrapolating the self-energy to temperatures below
$T_c$, as discussed above. We define an effective sp spectrum according to 
Eq.~(\ref{def:zk1}) and solve the gap equation (\ref{eq:gapbcs}) with this
effective sp spectrum.

\begin{figure}[bth]
\begin{center}
\includegraphics[origin=bl,width=7.0cm]{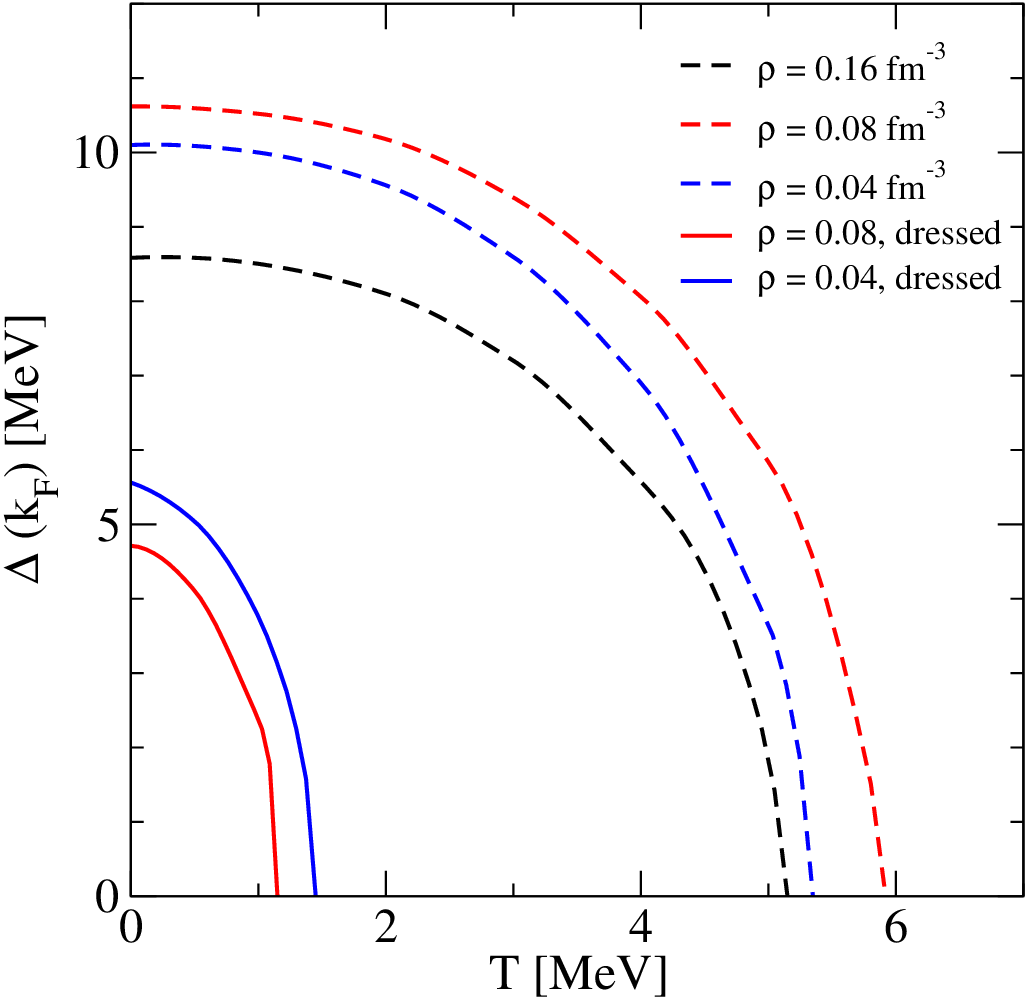}
\caption{Gap parameter $\Delta(k_F)$ in symmetric nuclear matter as a function
of temperature $T$. Results are presented for various densities, with and
without taking into account the dressing of the sp propagator due
to short-range correlations. The pairing gap disappears at $\rho$=0.16
fm$^{-3}$, if dressed propagators are considered.
All results displayed in this figure were obtained
using the CDBonn interaction.
\label{fig:gapnucl}}
\end{center}
\end{figure}

Results for the gap parameter $\Delta(k_F)$ in symmetric nuclear matter of
various densities are presented in Fig.~\ref{fig:gapnucl} as a function of the
temperature $T$. 
We will first discuss the results obtained within the usual BCS
approximation (see discussion above) in the $^3S_1-^3D_1$ partial wave. 
At the empirical saturation density the
CDBonn interaction yields a gap parameter $\Delta(k_F)$ at temperature 
$T=0$ of 
8.6 MeV (see above), which decreases with increasing temperature until it
vanishes at $T=5.2$ MeV. At $\rho$ = 0.08 fm$^{-3}$, which is about half the
empirical density, the value of the gap parameter at $T=0$ is even larger
($\Delta(k_F)$ = 10.6 MeV) and the gap calculated within the usual BCS approach
disappears only at a temperature of 5.9 MeV. This increase of the pairing gap
with decreasing density can be related to the momentum-dependence of the 
pairing
gap $\Delta (k)$ as displayed in Fig.~\ref{fig:gap3s2}: The gap function
increases with decreasing momentum. Therefore, as the Fermi momentum decreases
with density, the value $\Delta(k_F)$ tends to decrease with density. At even
lower densities, however, this effect is more than compensated by the feature,
that the phase-space of two-hole configurations decreases with density, so that
ultimately the gap parameter will approach the binding energy of the deuteron 
in the limit of $\rho \to 0$. 
This explains the decrease of the gap parameter going
from $\rho$ = 0.08 fm$^{-3}$  to  $\rho$ = 0.04 fm$^{-3}$. 

If we take the effects of short-range correlations into account, the 
generalized
gap equation of Eq.~(\ref{eq:gappw}) does not give a non-trivial solution for
symmetrical nuclear matter at $\rho$ = 0.16 fm$^{-3}$. This means that a proper
treatment of correlation effects in nuclear matter at normal density yields a
disappearance of the proton-neutron pairing predicted by the usual 
BCS approach.
The effects of short-range correlations tend to decrease with density. As a
consequence we obtain non-vanishing gaps for proton-neutron pairing at lower
densities (see solid lines in Fig.~\ref{fig:gapnucl}). This also leads to an 
increase of the critical temperature and the value of $\Delta(k_F)$ at $T=0$
going from $\rho$ = 0.08 fm$^{-3}$ to $\rho$ = 0.04 fm$^{-3}$. Note, that these
functions are qualitatively different from the corresponding BCS predictions.
\begin{figure}[bth]
\begin{center}
\includegraphics[origin=bl,width=7.0cm]{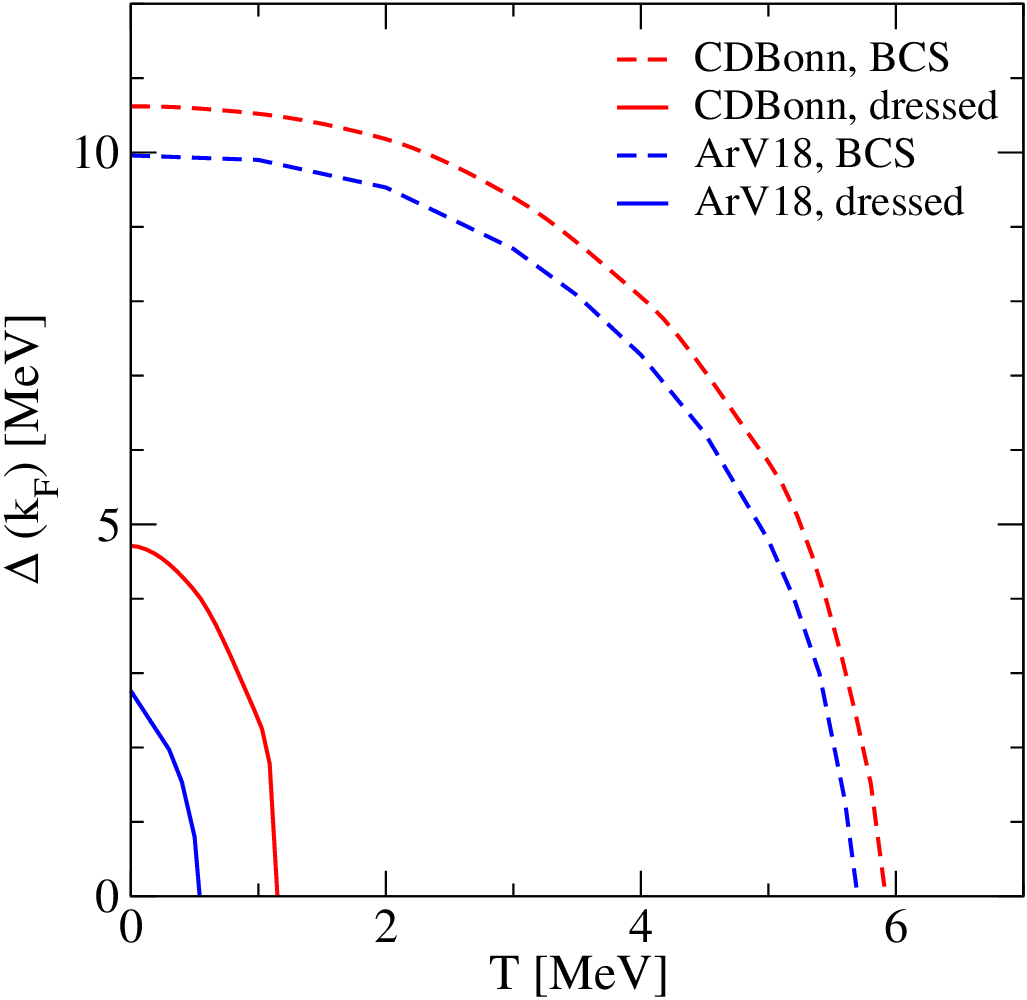}
\caption{Gap parameter $\Delta(k_F)$ in symmetric nuclear matter at $\rho$=0.08
fm$^{-3}$ as a function of temperature $T$ using the ArV18 and the CDBonn
interaction. 
\label{fig:gapnucl2}}
\end{center}
\end{figure}
Differences associated with the various interactions are displayed in
Fig.~\ref{fig:gapnucl2}. For nuclear matter with a density of $\rho$ = 0.08 
fm$^{-3}$ the gap parameter is presented as a function of temperature $T$ using
the BCS approximation and the generalized gap equation with dressed 
propagators.
As has already been discussed above, the ArV18 interaction yields smaller 
values
for the gap parameter and the critical temperature than the CDBonn interaction.

\begin{figure}[bth]
\begin{center}
\includegraphics[origin=bl,width=7.0cm]{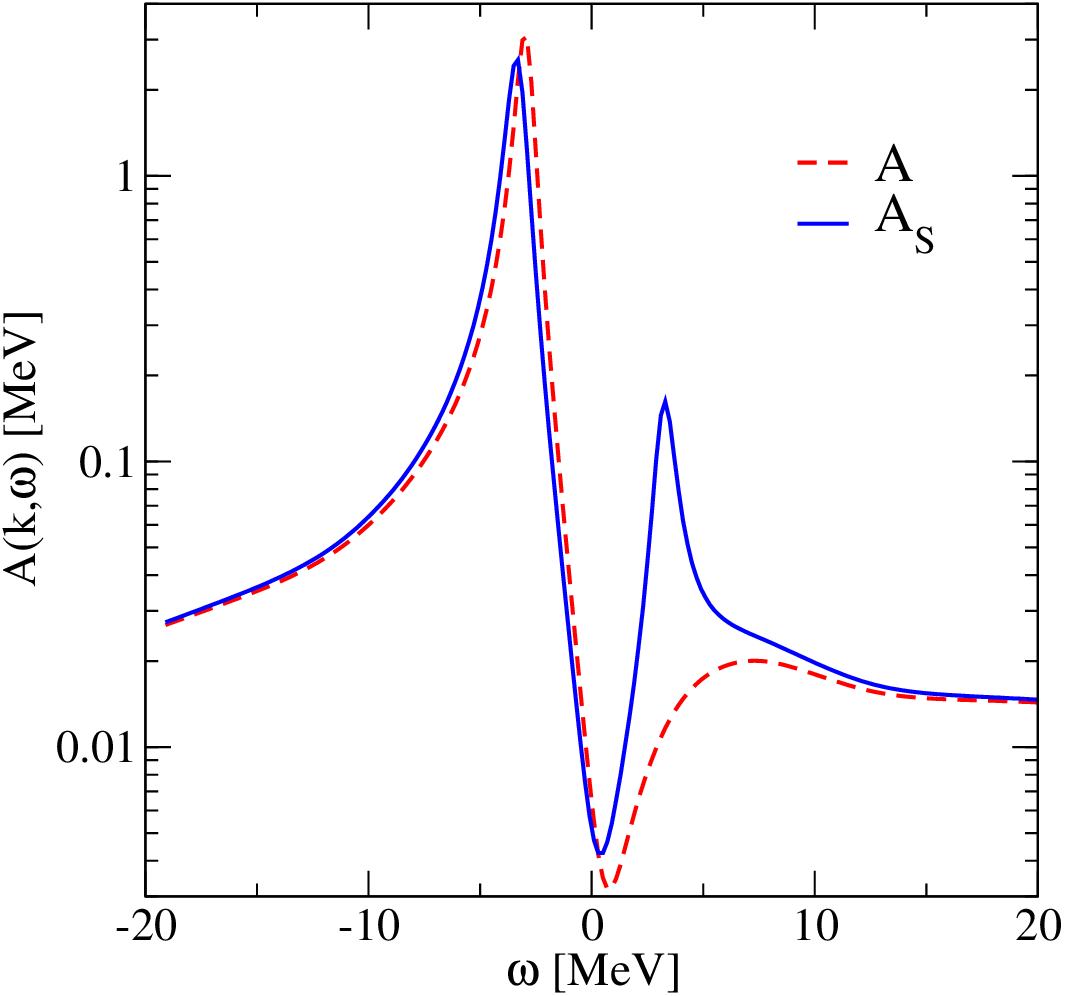}
\caption{Spectral function for nucleons with momentum $k$ = 193 MeV/c with
(solid line) and without (dashed line) inclusion of pairing correlations.
Results are presented for nuclear matter of $\rho$ = 0.08 fm$^{-3}$ at a
temperature $T$ = 0.5 MeV.
\label{fig:asm2}}
\end{center}
\end{figure}

Effects of pairing correlations on the spectral function are visualized in
Fig.~\ref{fig:asm2}. As an example we consider nuclear matter at $\rho$ = 0.08
fm$^{-3}$ and show results for the spectral function without 
($A(k,\omega)$) and
with inclusion of pairing correlations ($A_s(k,\omega)$, see
Eq.~(\ref{spec_g2s})). 
The momentum considered for this figure, $k$ = 193 MeV/c,
is slightly below the Fermi momentum $k_F$ = 208 MeV/c. One observes that the
inclusion of pairing correlations enhances the maximum of the spectral
distribution at positive values of $\omega$ considerably and shifts the
quasi-particle peak to more negative values of $\omega$. The pairing
correlations modify the spectral distribution into the direction which is
obtained in the simple BCS approximation for $A_s(k,\omega)$ in 
Eq.~(\ref{eq:specbcs}). Also note, that these modifications of the spectral
function $A_s(k,\omega)$ as compared to $A(k,\omega)$ is limited to a small
interval of energies around $\omega=0$ and to momenta close to the Fermi
momentum.

\begin{figure}[bth]
\begin{center}
\includegraphics[origin=bl,width=7.0cm]{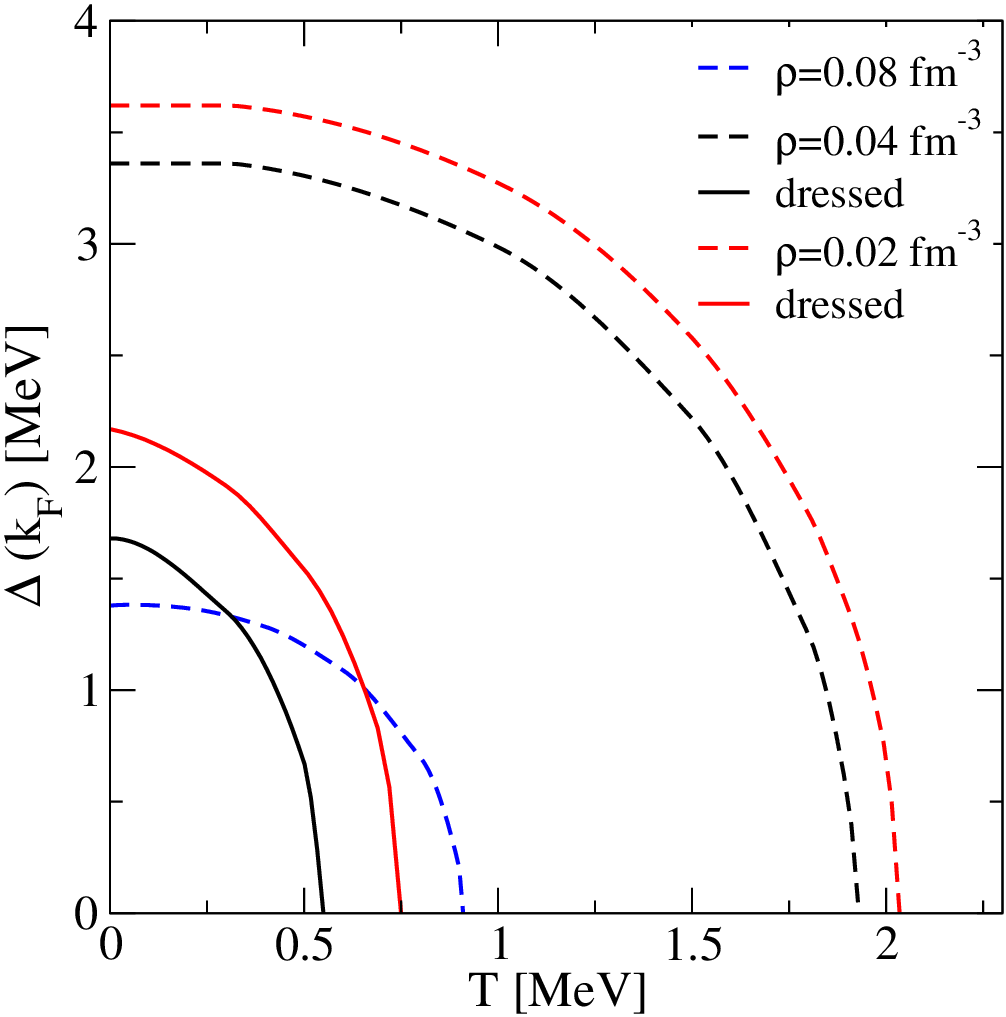}
\caption{Gap parameter $\Delta(k_F)$ in neutron matter as a function of 
temperature $T$. Results are presented for the usual BCS approximation (dashed
lines) and the solution of the generalized gap equation
(\protect{\ref{eq:gappw}}) in the $^1S_0$ partial wave using the CDBonn
interaction.
\label{fig:gapneutr}}
\end{center}
\end{figure}

Finally, we consider the case of neutron-neutron pairing in pure neutron 
matter.
We will focus the attention to densities, where the pairing correlations in the
$^1S_0$ partial wave are dominating. Results for the gap parameter $\Delta
(k_F)$ as a function of temperature are displayed in Fig.~\ref{fig:gapneutr}.
Using the BCS approximation with sp energies derived from the
quasi-particle energies of SCGF calculations we obtain a gap at $T=0$, which,
for the range of densities considered, increases with decreasing density. This
is in agreement with results of similar calculations, which are summarized 
\textit{e.g.} in \cite{dehj:03}. 

The effects of short-range correlations are weaker in neutron matter than in
nuclear matter. 
This has been discussed already above in connection with the results
displayed in Fig.~\ref{fig:compself}. This can also be seen in a comparison of
the dressed two-particle propagator and the effective strength factor $Z_{eff}$
defined in Eq.~(\ref{def:zk2}). While a calculation of $Z_{eff}$ for symmetric
nuclear matter at $\rho$ = 0.16 fm$^{-3}$ yields values for $Z_{eff}$, 
which are
typically around 0.8 (see Fig.~\ref{fig:zfact}), corresponding values for 
$Z_{eff}$ in neutron matter at $\rho$ = 0.08 fm$^{-3}$ are around 0.9.
Nevertheless, also these weaker effects of short-range correlations in neutron
matter are sufficient to suppress the formation of a pairing gap in neutron 
matter at $\rho$ = 0.08 fm$^{-3}$. 
Such a suppression of pairing correlations is also
observed at smaller densities. In this case, however, the inclusion of the
correlation effects just leads to a reduction of the gap parameter at a given
temperature and a reduction of the critical temperature (see
Fig.~\ref{fig:gapneutr}).  Extrapolating our results to neutron matter with
higher densities, we expect that the short-range correlations will suppress the
formation of pairing in the $^3P_2-^3F_2$ partial waves at those 
densities~\cite{khcl}.  

\section{CONCLUSIONS}
An attempt has been made to treat the effects of short-range and pairing
correlations in a consistent way within the $T$-matrix approach of the
self-consistent Green's function (SCGF) method. The pairing effects are
determined from a generalized gap equation that employs sp propagators
fully dressed by short-range and tensor correlations.
This equation is directly linked to the homogeneous solution of the $T$-matrix
equation of NN scattering in the medium, which is one of the basic
equations of the SCGF approach at temperatures above the critical temperature
for a phase transition to pairing condensation.
While short-range and tensor correlations yield a redistribution of
sp strength over a wide range of energies, the effects of pairing
correlations on the spectral function are limited in nuclear matter to a
relatively small interval in energy and momentum around the Fermi surface. 

The formation of a pairing gap is very sensitive to the quasi-particle energies
and strength distribution at the Fermi surface and can be suppressed by 
moderate temperatures. 
The formation of short-range correlations are sensitive to a
larger range of energies and momenta. So we observe, that the non-local CDBonn
interaction is softer with respect to the formation of short-range correlations
but yields larger pairing gaps compared to the local ArV18 model for the NN
interaction.

From this sensitivity to different areas in momentum and energy one may 
conclude
that the features of short-range correlations should be rather similar in
studies of nuclear matter and finite nuclei. The investigation of pairing
phenomena, however, is rather sensitive \textit{e.g.} 
to the energy spectrum around the
Fermi energy. Therefore the shell effects of finite nuclei may lead to quite
different results for pairing properties than corresponding studies in infinite
matter.

The redistribution of sp strength due to the short-range
correlations has a significant effect on the formation of a pairing gap. While
the usual BCS approach predicts a gap for proton-neutron pairing in nuclear
matter at saturation density as large as 8 MeV, the inclusion of short-range
correlations suppresses this gap completely. Correlation effects are weaker at
smaller densities, but still lead to a significant quenching of the 
proton-neutron
pairing gap and to a reduction of the critical temperature for the phase
transition.
Compared to symmetric nuclear matter correlation effects are weaker in neutron
matter. 
Nevertheless, the inclusion of correlations suppresses the formation of a
gap for neutron-neutron pairing at $\rho$ = 0.08 fm$^{-3}$ completely and 
yields a significant quenching at lower densities.

The effects of dressed sp propagator in the generalized gap
equation could be described in terms of an effective strength 
factor $Z_eff{k}$,
which has been considered in the literature 
before~\cite{boz:03,boz:00,blsz:02}.
Unfortunately, we have not been able to derive the value of this strength 
factor from bulk properties of the self-energy.

Although the effects of pairing correlations on the sp Green's
function is weak and limited to a small range in energy and momentum, these
modifications are very important to extend SCGF calculations to densities and
temperatures that suffer from the so-called pairing instability. The present
study is a first step towards a consistent treatment of pairing and short-range
correlations.  

\vspace{.5truecm}
This work is supported by the U.S. National Science Foundation under Grant
No. PHY-0140316 and the ``Landesforschungsschwerpunkt Quasiteilchen'' of the
state of Baden W\"urttemberg. H.M. would like for the hospitality of the
Department of Physics of Washington University in St. Louis, where a major part
of this work has been done.

\end{document}